\documentclass[%
 reprint,
amsmath,amssymb,
aps,
pra,
longbibliography,
]{revtex4-2}
\usepackage{graphicx}% Include figure files
\usepackage{amsmath}
\usepackage{xr}
\usepackage{qcircuit}
\usepackage{pifont}
\usepackage[makeroom]{cancel}
\usepackage{graphicx}% Include figure files

\usepackage{amsmath}
\usepackage{pifont}
\usepackage{amssymb}
\usepackage{float}
\usepackage{xcolor}
\usepackage{mathtools}
\definecolor{porange}{RGB}{231, 117, 0}
\definecolor{plightgrey}{RGB}{238, 238, 238}

\definecolor{pdenim}{RGB}{30, 144, 255}
\definecolor{pgrey}{RGB}{51, 51, 51}%%{160, 160, 160}32, 54, 77 8, 66, 102
\definecolor{plightdenim}{RGB}{42, 102, 143} 
\definecolor{greyblue}{RGB}{186, 199, 216}
\usepackage{xr}
\usepackage[makeroom]{cancel}
\usepackage{tikz}
\usetikzlibrary{spy}
\usepackage{pgfplots}
\pgfplotsset{compat = newest}
\usepackage{pgfplotstable}
\usetikzlibrary{positioning, decorations.pathreplacing, calc, intersections, pgfplots.groupplots,fadings}
\usetikzlibrary{arrows,intersections}
\usepackage{tikz-cd}
\usepackage{xcolor}
\usetikzlibrary{fadings}
\usepackage{physics}
\usepackage{verbatim}
\usepackage{siunitx}
\usetikzlibrary{matrix}
\usepackage{dcolumn}% Align table columns on decimal point
\usepackage{bm}% bold math
\usetikzlibrary{fadings,decorations.markings}
\pgfkeys{
  /pgf/arrow keys/.cd,
  pitch/.code={%
    \pgfmathsetmacro\pgfarrowpitch{#1}
    \pgfmathsetmacro\pgfarrowsinpitch{abs(sin(\pgfarrowpitch))}
    \pgfmathsetmacro\pgfarrowcospitch{abs(cos(\pgfarrowpitch))}
  },
}
\pgfplotsset{colormap={CM}{rgb(-500)=(0,0,1) color(0)=(black) rgb255(1500)=(238,140,238)}}
\pgfdeclarearrow{
  name = Cone,
  defaults = {       % inherit from Kite
    length     = +2.3pt +3.4,
    width'     = +0pt +0.2,
    line width = +0pt 1 1,
    pitch      = +0, % lie on screen
  },
  cache = false,     % no need cache
  setup code = {},   % so no need setup
  drawing code = {   % but still need math
    \pgfmathsetmacro\pgfarrowhalfwidth{.38\pgfarrowwidth}
    \pgfmathsetmacro\pgfarrowhalfwidthsin{\pgfarrowhalfwidth*\pgfarrowsinpitch}
    \pgfpathellipse{\pgfpointorigin}{\pgfqpoint{\pgfarrowhalfwidthsin pt}{0pt}}{\pgfqpoint{0pt}{\pgfarrowhalfwidth pt}}
    \pgfusepath{fill}
    \pgfmathsetmacro\pgfarrowlengthcos{\pgfarrowlength*\pgfarrowcospitch}
    \pgfmathparse{\pgfarrowlengthcos>\pgfarrowhalfwidthsin}
    \ifnum\pgfmathresult=1
      \pgfmathsetmacro\pgfarrowlengthtemp{\pgfarrowhalfwidthsin*\pgfarrowhalfwidthsin/\pgfarrowlengthcos}
      \pgfmathsetmacro\pgfarrowwidthtemp{\pgfarrowhalfwidth/\pgfarrowlengthcos*sqrt(\pgfarrowlengthcos*\pgfarrowlengthcos-\pgfarrowhalfwidthsin*\pgfarrowhalfwidthsin)}
      \pgfpathmoveto{\pgfqpoint{\pgfarrowlengthcos pt}{0pt}}
      \pgfpathlineto{\pgfqpoint{\pgfarrowlengthtemp pt}{ \pgfarrowwidthtemp pt}}
      \pgfpathlineto{\pgfqpoint{\pgfarrowlengthtemp pt}{-\pgfarrowwidthtemp pt}}
      \pgfpathclose
      \pgfusepath{fill}
    \fi
    \pgfpathmoveto{\pgfpointorigin}
  }
}
\usepackage{tikz}
\usepgflibrary{patterns}
\usetikzlibrary{shapes.geometric, arrows}
\usepackage{amssymb}
\usepackage{float}
\usepackage{xcolor}
\usepackage{mathtools}
\usepgfplotslibrary{colorbrewer}
\usetikzlibrary{3d,perspective}

\tikzset
{%
   pics/spiral/.style 2 args={code=%
   {
      \draw[pic actions] plot[domain=#1:#2,samples=1001]
        ({0.5*exp(-0.5*\x*\x)*cos(1040*\x)},\x,
         {exp(-1*\x*\x)*sin(1040*\x)});
   }},
}

\tikzset
{%
   pics/well/.style 2 args={code=%
   {
      \draw[pic actions] plot[domain=#1:#2,samples=1001]
        (\x,{2*\x*\x},
         {2*\x*\x});
   }},
}
\usepackage{tikz-cd}
\usepackage{soul}
\usepackage{MnSymbol}
\usetikzlibrary{calc}
\usetikzlibrary{fadings}
\usepackage{physics}
\usepackage{verbatim}
\usepackage{siunitx}
\usetikzlibrary{matrix}
\usepackage{dcolumn}% Align table columns on decimal point
\usepackage{bm}% bold math
\usetikzlibrary{decorations.markings}
\usepackage{tikz}
\usetikzlibrary{calc}
\usetikzlibrary{positioning}
\usepackage{comment}
\usepackage{pgfplots}

\usetikzlibrary[pgfplots.groupplots]

\AtBeginDocument{\RenewCommandCopy\qty\SI}
\pgfplotsset{compat=1.18}

\begin{document}

\preprint{APS/123-QED}

\title{Measuring Entanglement by Exploiting its Anti-symmetric Nature}
%anti-symmetric geometric description of entanglement enables efficient extraction
%Fermionic Entanglement Probes enabled by a Geometric Point of View
%Statement 1: The geometric point of view reveals an exclusively anti-symmetric description of entanglement
%Statement 2: Fermions are fundamentally anti-symmetric and enable the efficient extraction of entanglement from a geometric point of view
%Geometric Insights into Entanglement reveal an enable Efficient Fermionic Entanglement Probes
%Anti-Symmetric geometric description of Entanglement enables its Efficient Probe
%Geometric Insights into Entanglement: Anti-Symmetric Description enables Fermionic Probes
 %\\Shortcut to Entanglement Extraction enabled by its Anti-Symmetric Nature 
%Anti-symmetric Nature Enables Efficient Entanglement Extraction}
%Anti-symmetric Nature of Entanglement Enables its Efficient Extraction}% Force line breaks with \\

\author{Peyman Azodi$^{1}$}
\email{pazodi@princeton.edu}  
\author{Benjamin Lienhard$^{1,2}$}
\author{and Herschel A. Rabitz$^1$}
\affiliation{$^1$ Department of Chemistry, Princeton University, Princeton, NJ 08544, USA}%Lines break automatically or can be forced with \\
\affiliation{$^2$ Department of Electrical and Computer Engineering, Princeton University, Princeton, NJ 08544, USA}
%\email{pazodi@princeton.edu}
     

%\affiliation{%


%\affiliation{
%\affiliation{


\date{\today}% It is always \today, today,

\begin{abstract}
Despite significant progress in experimental quantum sciences, measuring entanglement entropy remains challenging. Through a geometric perspective, we reveal the intrinsic anti-symmetric nature of entanglement. We prove that most entanglement measures, such as von Neumann and R\'enyi entropies, can be expressed in terms of exterior products, which are fundamentally anti-symmetric. Leveraging this, we propose utilizing the anti-symmetric nature of fermions to measure entanglement entropy efficiently, offering a resource-efficient approach to probing bipartite entanglement.

% Despite advances in experimental quantum sciences, measuring entanglement entropy remains challenging. We introduce a geometric approach to entanglement, revealing that it can be harnessed for its measurement. We demonstrate that nearly all entanglement metrics, including the von Neumann and R\'enyi entropies, can be expressed via exterior products corresponding to specific volumes within Hilbert space. Moreover, we demonstrate that in fermions, inherent anti-symmetries can be utilized to encode these volumes. Thus, offering a novel method to measure bipartite entanglement. 

%\begin{description}
%\end{description}
\end{abstract}

\maketitle

%\tableofcontents

%\section{\label{sec:level1}First-level heading:\protect\\ The line
%break was forced \lowercase{via} \textbackslash\textbackslash}

\textit{Introduction.---}
The intertwined dynamics between quantum systems, inexplicable by classical physics, are called entanglement~\cite{EinsteinPR1935EPR}. Entanglement is a critical resource in quantum technologies, offering advantages in sensing, computing, and communication over their classical counterparts~\cite{RevModPhys.81.865,plenio2007introduction,Pirandola:20,ekert1991quantum}. In addition, entanglement has been used to elucidate fundamental principles of condensed matter physics, such as thermalization in isolated quantum systems~\cite{RevModPhys.82.277,kaufman2016quantum, RevModPhys.91.021001,PhysRevX.5.031032,PhysRevLett.96.110404, magnon, azodi2024slowgrowthentanglementlongrange}. 

\par The difficulty in classically simulating the dynamics of highly entangled quantum states has led to the development of quantum simulators~\cite{feynman2018simulating,cirac2012goals}. Nevertheless, directly measuring entanglement, such as the von Neumann entropy $\mathbf{S}_{\rm vN}(\rho)=-\Tr({\rho \ln{\rho}})$, where $\rho$ is the density matrix, remains a cumbersome task~\cite{amico2008entanglement}. Early proposals aimed at reducing the experimental complexity, particularly for bosonic systems, focused on the second-order R\'enyi entropy $\mathbf{H}^{(\alpha)}(\rho)=\frac{1}{1-\alpha} \ln \Tr ({{\rho}}^\alpha)$ with $\alpha$ representing the order~\cite{mintert2005concurrence,zyczkowski2003renyi}. The second-order R\'enyi entropy can be extracted either through many-body interference between identical copies of the system~\cite{PhysRevLett.93.110501,daley2012measuring,PhysRevA.72.042335,islam2015measuring,doi:10.1126/science.aau0818,rispoli2019quantum} or via randomized measurement techniques~\cite{elben2018renyi,doi:10.1126/science.aau4963,elben2023randomized,imai2021bound,rath2021importance}. Recently, the classical shadow formalism has been proposed to obtain the moments $\Tr\left(\rho^k\right)$~\cite{huang2020predicting,neven2021symmetry}. 

\par While a complete characterization of a quantum state through tomography enables inferring the degree of entanglement, it becomes extremely resource-intensive as quantum systems scale in size~\cite{BanaszekIOP2013Tomo}. Entanglement witnesses offer an elegant solution to circumvent these substantial resource demands~\cite{bae2015quantum, girolami2017witnessing, brandao2005quantifying}. Utilizing the inherent symmetries within the quantum system as entanglement witnesses has proven effective, particularly in methods involving many-body interference among identical replicas of the quantum system~\cite{PhysRevLett.93.110501,daley2012measuring, beckey2021computable,PhysRevLett.125.180402}. 

\par In this manuscript, we focus on the intrinsic symmetries of entanglement. We prove that entanglement is fundamentally an anti-symmetric geometric feature of a system's state in Hilbert space. %However, most entanglement measures are algebraically symmetric, meaning they are unaffected by permuting the eigenvalues of the density matrix~\cite{RevModPhys.81.865}. The connection between these two aspects of entanglement forms the foundation of the presented proof. 
We show that the building blocks of bipartite entanglement are a set of anti-symmetric geometric structures. They are described via exterior products in the Hilbert space, which we refer to as higher-order volumes. We prove that these higher-order volumes form a basis for the space of symmetric measures of bipartite entanglement. Here, symmetric measures refer to invariant functions upon permuting the eigenvalues of the density matrix~\cite{RevModPhys.81.865}. Hence, a wide range of entanglement entropy measures can be uniquely described using these volumes up to a sufficient order.

\par The anti-symmetric character of entanglement allows for employing inherent symmetries of fermions to encode entanglement. Due to intrinsic fermionic symmetry rules, accumulation of fermions in the same state upon interference is prohibited~\cite{bocquillon2013coherence, Lim_2005}. However, we show that the probability of such occurrences for entangled fermions equals the higher-order volumes of interest. Therefore, we uncover a new link between entanglement geometry and inherent particle symmetries that can be leveraged to effectively extract entanglement.

\par In summary, we show that entanglement can be quantified by solely focusing on its anti-symmetric characteristics in Hilbert space. This approach can potentially streamline the process by eliminating the requirement for extensive post-processing calculations to derive these anti-symmetries, thus directly determining various entanglement entropies.

\par \textit{Geometrical description of entanglement entropy.---}
%Here, we demonstrate that entanglement between two parts of a pure quantum system can be quantified using anti-symmetric exterior tensor products. 
Consider a pure bipartite quantum system described by the state:
\begin{equation}\label{decom}
    \ket{\psi} =\sum_{{j},i} \psi_{{j},i} \ket{{j}}\otimes \ket{i},
\end{equation}
where $\{\ket{{j}}: j=1,\cdots,n\}$ and $\{\ket{{i}}:i=1,\cdots,d\}$ represent arbitrary basis states for an $n$-dimensional subsystem $\mathcal{M}$ and a $d$-dimensional accompanying subsystem $\mathcal{R}$, respectively. The projected and not necessarily normalized states of subsystem $\mathcal{R}$ ($\mathcal{M}$) are defined as 
\begin{equation}\label{expansion}
    \begin{split}
        \ket{{j} \psi}\doteq\left(\bra{{j}}\otimes I_{\mathcal{R}}\right) \ket{\psi}= \sum_{i} \psi_{{j},{i}}\ket{i}\\
        \left(\ket{{i} \psi}\doteq\left(I_{\mathcal{M}}\otimes \bra{{i}}\right) \ket{\psi}= \sum_{j} \psi_{{j},{i}}\ket{j}\right),
    \end{split}  
\end{equation} 
where $I_{\mathcal{R}}$ ($I_{\mathcal{M}}$) is the identity operator acting on subsystem ${\mathcal{R}}$ (${\mathcal{M}}$). 

\par The second-order R\'enyi entropy of the reduced density matrix $\rho_{\mathcal{M}}=\Tr_{\mathcal{R}}\{\ket{\psi}\bra{\psi}\}=\sum_{i=1}^d\ket{i\psi}\bra{i\psi}$ can be expressed as the linear entropy ($\mathbf{S}_{\rm L}$)~\cite{zyczkowski2003renyi}
\begin{equation}\label{lineare}
    \begin{split}
    \mathbf{S}_{\rm L}(\rho_{\mathcal{M}})&=1-e^{-\mathbf{H}^{(2)}(\rho_{\mathcal{M}})}=1-\Tr (\rho_{\mathcal{M}}^2)=1-\sum_{j=1}^n \lambda_j^2\\
    &=1-\left\{\left(\sum_{j=1}^n \lambda_j\right)^2-2\sum_{1\leq  {j}_{1}< {j}_2\leq n}\lambda_{j_1}\lambda_{j_2}\right\}\\
    &=2\sum_{1\leq  {j}_{1}< {j}_2\leq n}\lambda_{j_1}\lambda_{j_2},
    \end{split}
\end{equation}
where $\lambda_1,\dots,\lambda_n$ are the eigenvalues of $\mathcal{M}$.% and $\gamma=1-\mathbf{S}_{\rm L}$ represents the quantum purity.

\par The sum of products of all pairs of distinct eigenvalues in Eq.~(\ref{lineare}) is equal to the sum of all second principal minors of $\rho_{\mathcal{M}}$~\cite{meyer2023matrix}. Thus, a measure for entanglement between $\mathcal{M}$ and $\mathcal{R}$, monotonically related to the second-order R\'enyi and linear entropies~\cite{QCTF}, can be expressed as:
\begin{align}
    \begin{split}\label{qt}
        \tilde{\mathcal{Q}}_{\mathcal{M}}^{(2)}&=\frac{1}{2}\mathbf{S}_{\rm L}(\rho_{\mathcal{M}})= \sum_{1\leq  {j}_{1}< {j}_2\leq n}\lambda_{j_1}\lambda_{j_2},\\
        &=\sum_{1\leq  {j}_{1}< {j}_2\leq n} \left| \begin{array}{cc}
        \bra{j_1}\rho_{\mathcal{M}}\ket{j_1} & \bra{j_1}\rho_{\mathcal{M}}\ket{j_2} \\
        \bra{j_2}\rho_{\mathcal{M}}\ket{j_1} & \bra{j_2}\rho_{\mathcal{M}}\ket{j_2} 
        \end{array} \right|\\
        &=\sum_{1\leq  {j}_{1}< {j}_2\leq n} \left| \begin{array}{cc}
        \braket{j_1\psi}{j_1\psi} & \braket{j_1\psi}{j_2\psi} \\
        \braket{j_2\psi}{j_1\psi} & \braket{j_2\psi}{j_2\psi} 
        \end{array} \right|\\
        & = \sum_{1\leq  {j}_{1}< {j}_2\leq n} {\left(\bra{{j}_1 \psi} \wedge  \bra{{j}_2 \psi})(\ket{{j}_1 \psi} \otimes \ket{{j}_2 \psi}\right)},
    \end{split}
\end{align}
where the wedge operator $\wedge$ denotes the anti-symmetric exterior tensor product which is defined to second order as:
\begin{equation}\label{wedgee}
    \begin{split}
        \bra{{j}_1\psi}\wedge\bra{{j}_2\psi}&=-\bra{{j}_2\psi}\wedge\bra{{j}_1\psi}\\
        &\coloneq\bra{{j}_1\psi}\otimes\bra{{j}_2\psi}-\bra{{j}_2\psi}\otimes \bra{{j}_1\psi}.
    \end{split}
\end{equation}

\par Eq.~(\ref{qt}) can be generalized to higher orders utilizing the wedge product $r$-th order:
\begin{equation}\label{wedge}
    \bra{{{j}_1} \psi} \wedge :\wedge \bra{{{j}_r} \psi}\doteq 
    \sum_{\{\pi\}} {\epsilon (\pi) \bra{{{j}_{\pi (1)}} \psi} \otimes  :\otimes \bra{{{j}_{\pi(r)}} \psi}},
\end{equation}
summing over the set $\{\pi\}$ of all permutations of $\{1,2\dots,r\}$ with $\epsilon(\pi)$ representing the sign of each permutation. 

\par The wedge product between $r$ covariant (dual or bra) vectors in Eq.~(\ref{wedge}) is called an $r-$form. The inner product between $r$-forms, denoted by $\nu=\bra{\nu_1} \wedge \cdots\wedge \bra{\nu_r }$ and  $\mu=\bra{\mu_1 } \wedge\cdots\wedge \bra{\mu_r }$, is given by the Gram determinant 
\begin{equation}\label{EQ:gram}\braket{\nu}{\mu}_{\rm G}=\det \{(\braket{\nu_i}{\mu_j})_{i,j=1,\cdots,r}\}.
\end{equation}
Accordingly, one can define the norm $|\nu|^2=\braket{\nu}{\nu}_{\rm G}$. Using this definition, the argument in the summation in Eq.~(\ref{qt}) can be expressed as $|\bra{{j}_1 \psi} \wedge  \bra{{j}_2 \psi}|^2$.

%\par The argument in the summation in Eq.~(\ref{qt}) represents the squared area (equivalent to two-dimensional volume) spanned by the corresponding projected wave functions. Furthermore, it can be expressed as $|\bra{{j}_1 \psi} \wedge  \bra{{j}_2 \psi}|^2$. 
%%This quantity captures second-order correlations in Hilbert space~\cite{QCTF}. 

\par \textit{Higher-order correlations and elementary symmetric polynomials.---}Higher-order correlations between subsystems can be determined by considering exterior products between more than two covectors of projected states. The equation
\begin{equation}\label{vol}
    |v_r|^2\doteq\!\sum_{1\leq  {j}_1<:< {j}_r\leq n}\!{\left(\bra{{{j}_1} \psi} \wedge :\wedge \bra{{{j}_r} \psi})(\ket{{{j}_1} \psi} \otimes :\otimes \ket{{{j}_r} \psi}\right)}
\end{equation}
yields the total collective squared $r$-th-order volume spanned by all $r$ projected states in Hilbert space. This measure, $|v_r|^2$, quantifies the pure $r$-th-order correlation between two subsystems within Hilbert space, as will be shown next.
\par The general term $r$-th-order volume is a measure of the space enclosed within $r$ vectors. For example, the $2$-nd order volume is the area enclosed by the parallelogram created by two vectors, and the $3$-rd-order volume is the volume of the parallelotope created by three vectors. Analogously, the $r$-th-order volume is the hyper-volume created using $r$ vectors. The definition of $r$-th-order volumes through wedge products is sensitive to the permutation of the vector. For example, in Eq.~(\ref{wedgee}) by exchanging $\bra{j_1\psi}$ and $\bra{j_2 \psi}$, a negative sign is introduced. However, the definition in Eq.~(\ref{vol}) is the norm of the $r$-th-order volume and invariant under exchanging the vectors.

\par Using the inner product relation between the $r$-forms, each summand in Eq.~(\ref{vol}) can be expressed as the following determinant
\begin{equation}\label{minor}
    \begin{vmatrix}
\braket{{j}_1\psi}{{j}_1\psi} & \braket{{j}_1\psi}{{j}_2\psi}& \cdots&\braket{{j}_1\psi}{{j}_r\psi}\\
\vdots &\vdots& \ddots & \vdots
\\
\braket{{j}_r\psi}{{j}_1\psi} & \braket{{j}_r\psi}{{j}_2\psi} & \cdots &\braket{{j}_r\psi}{{j}_r\psi},
\end{vmatrix}.
\end{equation}
which is a principal minor of the reduced density matrix $\rho_{\mathcal{M}}$. Hence, Eq.~(\ref{vol}) is the summation of all possible principal minors of the reduced density matrix. The reduced density matrix is linked to the eigenvalues of the density matrix through a fundamental property of square matrices; the summand in Eq.~(\ref{vol}) equals the so-called elementary symmetric polynomial (ESP) in $r$ eigenstates $\{\lambda_1, \cdots ,\lambda_n\}$ of $\rho_{\mathcal{M}}$ defined as:
\begin{equation} \label{ESP}
    e_r=\sum _{1\leq k_1 <k_2<\cdots<k_r\leq n} \lambda_{k_1}\lambda_{k_2} \cdots \lambda_{k_r}. 
\end{equation}
This implies that $|v_r|^2=e_r$ and is proven  in Ref.~\cite{meyer2023matrix}. 

\par The ESPs are algebraically independent and span the space of symmetric polynomials of eigenvalues. Symmetric polynomials are invariant under the permutations of their arguments. Therefore, all entanglement entropy measures that can be expressed as a symmetric polynomial of eigenvalues, including the R\'enyi and von Neumann entropies, can be uniquely represented via ESPs. Eqs.~(\ref{minor}) and (\ref{ESP}) capture the transition and link between the squared anti-symmetric geometric features and algebraically symmetric fundamental expressions, respectively. This link enables the description of any measures of entropy in terms of purely anti-symmetric geometric structures.%

\par \textit{Bipartite entanglement measure representations through ESPs.---}%The possibility of representing any symmetric measure of bipartite entanglement through higher-order volumes, $|v_r|^2$, or equivalently, ESPs, $e_r$, is the direct application of . 
A symmetric polynomial of order $r$, such as the measures of bipartite entanglement, can be uniquely represented via ESPs $e_1, e_2, \cdots, e_r$~\cite{macdonald1998symmetric, tignol2015galois}. The order of a symmetric polynomial is the maximum sum of exponents among terms involving the eigenvalues. To show this phenomenon, we will discuss such representations for the von Neumann entropy and the family of higher-order purities $\Tr\left(\rho_{\mathcal{M}} ^k\right)$ through ESPs. 

\par Higher-order purities in the R\'enyi entropy family can be expressed using the Girard–Newton formula~\cite{tignol2015galois}:
\begin{equation}\label{negativities}
    \Tr\left(\rho_\mathcal{M} ^k\right)= k (-1)^k \sum '  \left\{\left(\sum_{l=1}^{k} p_l -1\right)! \prod _{l=1}^{k} \frac{{(-e_l)}^{p_l}}{p_l !} \right\},
\end{equation}
where the primed summation is over all possible $k$-tuples $(p_1,\cdots,p_k)$ with $k\in \mathbb{Z}^+$ such that $\sum_{l=1}^k lp_l=k$ with $p_1\geq 0, \dots, p_k\geq 0$. Since $e_k=0$ for $k>n$, and given the first $n$ ESPs, one can obtain the countably infinite family of the higher-order purities that follow the hierarchy of R\'enyi and von Neumann entropies. Note that the inner summation in Eq.~(\ref{negativities}) can be written in terms of the ordinary Bell polynomials~\cite{macdonald1998symmetric}, which allows for a simple computation of the higher-order purities. 

\par The von Neumann entropy $\textbf{S}_{\rm vN}(\rho_{\mathcal{M}})$ can be uniquely characterized by ESPs if all ESPs up to order $n$ are included. For instance, the von Neumann entropy $\textbf{S}_{\rm vN}(\rho_{\mathcal{M}})$, after a Taylor expansion, usage of the binomial expansion, and Eq.~(\ref{negativities}), can be expressed as:
\begin{equation}\label{pur}
    \begin{split}
        \textbf{S}_{\rm vN}(\rho_{\mathcal{M}})=&-\sum_{m=1}^\infty\frac{1}{m}\sum_{k=1}^{m+1}k\binom{m}{k-1} \\& \sum_{\substack{\sum_{l=1}^n lp_l=k\\p_1\geq 0, \dots, p_n\geq 0}}\left(\sum_{l=1}^{n} p_l -1\right)!\prod _{l=1}^{n} \frac{{(-e_l)}^{p_l}}{p_l !}.
    \end{split}
\end{equation}

Moreover, the following formulation introduces an approximation of the von Neumann entropy as a function of the first $r <n$ EPSs. In the Supplemental Material~\cite{supp}, we show that using Eq.~(\ref{pur}), after truncating to $r$-th order in purities, the von Neumann entropy is approximated by the following $r$-th-order entanglement entropy $\mathbf{S}_r$ as:
\begin{widetext}
\begin{equation}\label{EEf}
  \begin{split}
    \mathbf{S}{\rm vN}(\rho_{\mathcal{M}})\gtrsim\mathbf{S}_{r}(\rho_{\mathcal{M}})&=-\sum_{m=1}^{\infty }\frac{1}{m} \sum_{\substack{0<\sum_{l=1}^r lp_l\leq m+1\\p_2\geq 0, \dots, p_r\geq 0}} (-1)^{\sum^r_{l=2} (l-1)p_l} \prod_{l=2}^r \frac{e_l^{p_l}}{p_l!}\sum_{k=\sum^r_{l=1}lp_l}^{m+1} k (-1)^k \binom{m}{k-1}\frac{\left(k-1-\sum^r_{l=2}(l-1)p_l\right)!}{\left(k-\sum^r_{l=2}lp_l\right)!}.
  \end{split}
\end{equation}  
\end{widetext}

\par This finite series of entanglement entropy measures approaches the von Neumann entropy $\mathbf{S}_{n}=\mathbf{S}_{\rm vN}$. When $i\ll n$, the series of $e_i$ decays faster than $1/{i!}$. Therefore, the lower orders of ESPs can efficiently approximate the value of the purities and the von Neumann entropy, as numerically illustrated in the Supplemental Material~\cite{supp}.

\par \textit{Measuring $|v_r|^2$ through fermionic anti-symmetries.---}Consider a closed quantum system that includes a subsystem $\mathcal{M}$ comprising a single fermion. The fermion's internal energy level $j$ is created by the operator $a^\dagger _j$. Using the decomposition introduced in Eqs.~(\ref{decom}) and (\ref{expansion}), the state of the closed system can be written as
\begin{equation}
    \ket{\psi}=\sum_j a^\dagger_j\ket{0}\otimes \ket{j\psi}.
\end{equation}
Accordingly, by preparing two copies of the system, the joint state is given by
\begin{equation}\label{joint}
    \sum_{j_1,j_2} {{a^\dagger}_{j_1}}^{(1)}{{a^\dagger}_{j_2}}^{(2)}\ket{0,0}_{\rm{in}}\otimes \ket{j_1\psi}\otimes \ket{j_2\psi},
    % \sum_{j_1,j_2} {{a^\dagger}_{j_1}}^{(1)}\ket{0}^{(1)}\otimes{{a^\dagger}_{j_2}}^{(2)}\ket{0}^{(2)}\otimes \ket{j_1\psi}\otimes \ket{j_2\psi},
\end{equation}
where the superscripts $(1)$ and $(2)$ distinguish the different copies of the system. Inspired by prior interference experiments among un-entangled fermions~\cite{neder2007interference, oliver1999hanbury, ji2003electronic, bocquillon2013coherence, preiss2019high, PhysRevLett.96.080402}, the fermions $(1)$ and $(2)$ are sent to separate input ports of a beamsplitter, denoted by the $\ket{}_{\rm{in}}$. %\textcolor{red}{\st{with the input ports indicated as superscripts $(1)$ and $(2)$.}} \textcolor{red}{\st{Note that the input ports are initialized as $\ket{0}^{(c)}$ with $c={1,2}$.}}
A $50:50$ beamsplitter transfers the creation operators at the input ports ${a^\dagger_j}^{(c)},c=1,2$ to the creation operators at the output ports ${b^\dagger_j}^{(c)}, c={3,4}$ as
\begin{equation}\label{interf}
    \begin{split}
    {a^\dagger_j}^{(1)}\rightarrow \frac{1}{\sqrt{2}}\left({b^\dagger_j}^{(3)}-{b^\dagger_j}^{(4)}\right),\\{a^\dagger_j}^{(2)}\rightarrow \frac{1}{\sqrt{2}}\left({b^\dagger_j}^{(3)}+{b^\dagger_j}^{(4)}\right).
    \end{split}
\end{equation}

\par The probability of detecting fermions at the same output ---fermion bunching--- is determined by the component of the state comprising terms of the form ${{b^\dagger}_{j_1}}^{(c)}{{b^\dagger}_{j_2}}^{(c)}$. This part of the state can be obtained by substituting Eq.~(\ref{interf}) in Eq.~(\ref{joint}), and keeping the terms involving ${{b^\dagger}_{j_1}}^{(c)}{{b^\dagger}_{j_2}}^{(c)}$:
\begin{equation}\label{stat}\begin{split}
    \frac{1}{2} \sum _{j_1, j_2} &\big({{b^\dagger}_{j_1}}^{(3)}{{b^\dagger}_{j_2}}^{(3)}- {{b^\dagger}_{j_1}}^{(4)}{{b^\dagger}_{j_2}}^{(4)}\big )\ket{0,0}_{\rm{out}}\otimes \ket{j_1\psi}\otimes \ket{j_2\psi}
     %\frac{1}{2} \sum _{j_1, j_2} &\left({{b^\dagger}_{j_1}}^{(3)}{{b^\dagger}_{j_2}}^{(3)}\ket{0}^{(3)}\otimes\ket{0}^{(4)}\right.
    %\\
    %&\left.- \ket{0}^{(3)}\otimes{{b^\dagger}_{j_1}}^{(4)}{{b^\dagger}_{j_2}}^{(4)}\ket{0}^{(4)}\right)\otimes \ket{j_1\psi}\otimes \ket{j_2\psi}.
\end{split}
\end{equation}
The creation operators obey the anti-commutation relations $\left\{{b^{\dagger}_{j_1}}^{(c%_1
)}, {b^{\dagger}_{j_2}}^{(c%_2
)}\right\}=0$, for %$c_1,c_2=3,4$
$c={3,4}$. A resultant property ${b^\dagger_{j}}^{(c)}{b^\dagger_{j}}^{(c)}=0$ can be used to exclude the terms with $j_1=j_2$ in the summation. Moreover, we have ${b^\dagger_{j_1}}^{(c)}{b^\dagger_{j_2}}^{(c)}=-{b^\dagger_{j_2}}^{(c)}{b^\dagger_{j_1}}^{(c)}$. Using this property, the summation can be rewritten over the pairs of $j_1<j_2$, instead of the original summation over $j_1,j_2$. Therefore, Eq.~(\ref{stat}) is
 \begin{equation}\label{stat3}\begin{split}
     \frac{1}{2} \sum _{j_1< j_2} &{{b^\dagger}_{j_1}}^{(3)}{{b^\dagger}_{j_2}}^{(3)}\ket{0,0}_{\rm{out}}\otimes \ket{j_1\psi}\otimes \ket{j_2\psi}
     \\ -&{{b^\dagger}_{j_1}}^{(3)}{{b^\dagger}_{j_2}}^{(3)}\ket{0,0}_{\rm{out}}\otimes \ket{j_2\psi}\otimes \ket{j_1\psi}
     \\-& {{b^\dagger}_{j_1}}^{(4)}{{b^\dagger}_{j_2}}^{(4)}\ket{0,0}_{\rm{out}}\otimes \ket{j_1\psi}\otimes \ket{j_2\psi}
 \\+& {{b^\dagger}_{j_1}}^{(4)}{{b^\dagger}_{j_2}}^{(4)}\ket{0,0}_{\rm{out}}\otimes \ket{j_2\psi}\otimes \ket{j_1\psi},
 \end{split}
 \end{equation}
%\begin{equation}\label{stat3}\begin{split}
%    \frac{1}{2} \sum _{j_1< j_2} &{{b^\dagger}_{j_1}}^{(3)}{{b^\dagger}_{j_2}}^{(3)}\ket{0}^{(3)}\otimes\ket{0}^{(4)}\otimes \ket{j_1\psi}\otimes \ket{j_2\psi}
%    \\ -&{{b^\dagger}_{j_1}}^{(3)}{{b^\dagger}_{j_2}}^{(3)}\ket{0}^{(3)}\otimes\ket{0}^{(4)}\otimes \ket{j_2\psi}\otimes \ket{j_1\psi}
 %   \\-&\ket{0}^{(3)}\otimes{{b^\dagger}_{j_1}}^{(4)}{{b^\dagger}_{j_2}}^{(2)}\ket{0}^{(4)}\otimes \ket{j_1\psi}\otimes \ket{j_2\psi}
%\\+& \ket{0}^{(3)}\otimes{{b^\dagger}_{j_1}}^{(4)}{{b^\dagger}_{j_2}}^{(2)}\ket{0}^{(4)}\otimes \ket{j_2\psi}\otimes \ket{j_1\psi},
%\end{split}
%\end{equation}
or equivalently,
% \begin{equation}\label{stat2}\begin{split}
%     \sum _{j_1< j_2} &\Big ({{b^\dagger}_{j_1}}^{(1)}{{b^\dagger}_{j_2}}^{(1)}-{{b^\dagger}_{j_1}}^{(2)}{{b^\dagger}_{j_2}}^{(2)}\big )\ket{0,0}_{\rm{out}}\\&\otimes \frac{1}{2} \big(\ket{j_1\psi}\otimes  \ket{j_2\psi}-\ket{j_2\psi}\otimes \ket{j_1\psi} \big )
% \end{split}
% \end{equation}

\begin{equation}\label{stat2}\begin{split}
    \sum _{j_1< j_2} &\left({{b^\dagger}_{j_1}}^{(3)}{{b^\dagger}_{j_2}}^{(3)}\ket{0,0}_{\rm{out}}-{{b^\dagger}_{j_1}}^{(4)}{{b^\dagger}_{j_2}}^{(4)}\ket{0,0}_{\rm{out}}\right)
    \\&\otimes \frac{1}{2} \big(\ket{j_1\psi}\otimes  \ket{j_2\psi}-\ket{j_2\psi}\otimes \ket{j_1\psi} \big ).
\end{split}
\end{equation}
%\begin{equation}\label{stat2}\begin{split}
%    \sum _{j_1< j_2} &\left({{b^\dagger}_{j_1}}^{(3)}{{b^\dagger}_{j_2}}^{(3)}\ket{0}^{(3)}\otimes\ket{0}^{(4)}\right.
%    \\&\left.-\ket{0}^{(3)}\otimes{{b^\dagger}_{j_1}}^{(4)}{{b^\dagger}_{j_2}}^{(2)}\ket{0}^{(4)}\right)
%    \\&\otimes \frac{1}{2} \big(\ket{j_1\psi}\otimes  \ket{j_2\psi}-\ket{j_2\psi}\otimes \ket{j_1\psi} \big )
%\end{split}
%\end{equation}
Hence, for each pair ($j_1<j_2$), the probability of finding both fermions at the same output port of the beamsplitter is given by the norm of the term on the last line of Eq.~(\ref{stat2}) times $2$, to account for both outputs. Therefore, the fermion bunching probability follows as
\begin{equation}\begin{split}
&\mathbf{P}_{\text{Fermion bunching}}\\&=\frac{2}{4}\sum_{j_1<j_2} |\ket{j_1\psi}\otimes  \ket{j_2\psi}-\ket{j_2\psi}\otimes \ket{j_1\psi} |^2 
 \\&=\frac{2}{4} \times 2 \times \sum_{j_1<j_2}\braket{j_1\psi}{j_1\psi}\braket{j_2\psi}{j_2\psi}-\braket{j_1\psi}{j_2\psi}\braket{j_2\psi}{j_1\psi}
 \\&=\sum_{  {j}_{1}< {j}_2} \left| \begin{array}{cc}
        \braket{j_1\psi}{j_1\psi} & \braket{j_1\psi}{j_2\psi} \\
        \braket{j_2\psi}{j_1\psi} & \braket{j_2\psi}{j_2\psi} 
        \end{array} \right|
\\&=|v_2|^2,
\end{split}    
\end{equation}
using the definition of the $r-$th-order volume introduced in Eqs.~(\ref{vol}) and (\ref{minor}).

\par Fermion bunching is forbidden for unentangled fermions~\cite{bocquillon2013coherence, neumann2010single}. However, we showed that the anti-symmetrization is altered for fermions entangled to the accompanying subsystem, $\mathcal{R}$. This phenomenon leads to a fundamental link between inherent fermionic anti-symmetries and a unique geometric characterization of entanglement. In the Supplemental Material~\cite{supp}, we show that the described method can be generalized to encode the $r$'th order volume $|v_r|^2$ by employing $r-1$ beamsplitters across $r$ replicas of the system.

\par \textit{Concluding Remarks---}In this manuscript, we put forth a new geometric paradigm to characterize entanglement in quantum systems to address the tantalizing challenge of efficient entanglement extraction~\cite{200502, pichler2016measurement,li2008entanglement, elben2020mixed, leone2022stabilizer,oliviero2022measuring, liu2022detecting, gray2018machine}. The proposed formulation explicitly translates bipartite entanglement in many-body quantum systems into purely anti-symmetric geometric variables within the Hilbert space. These variables can be directly linked to fermionic symmetries, providing a resource-efficient technique to extract entanglement.

\par The introduced geometric characteristics of entanglement can potentially be leveraged to extract entanglement metrics across various quantum platforms, such as cold atoms. In these setups, all the necessary components for measuring higher-order volumes in multi-level fermionic systems are available~\cite{schreiber2015observation, guardado2021quench, mazurenko2017cold}. Specifically, the proposed systematic approach to probing the evolution of von Neumann entropy in quench experiments enables the exploration of fundamental aspects of out-of-equilibrium quantum behavior.We proved that the higher-order volumes form a basis for the space of symmetric entanglement measures. Two consequences of this statement are: First, any symmetric measure of entanglement can be completely and uniquely represented using this basis. Second, this basis can link various notions of entanglement entropy, including the von Neumann, R\'enyi, and the introduced $\mathbf{S}_r$ entropies. Moreover, the geometric anti-symmetric nature of this basis enables the direct encoding into collective particles' symmetries. This connection can lead to a direct encoding of entanglement into many-body symmetries. We show this connection in the context of fermionic systems. By providing the geometric perspective, our theoretical study complements the existing research on the link between entanglement and particle symmetries, which have examined both bosons~\cite{daley2012measuring, kaufman2016quantum} and fermions~\cite{PhysRevLett.125.180402}.

\begin{acknowledgments}
P.A. expresses gratitude to Lily Li for her insightful comments on the algebraic independence of ESPs and Ian Fleschler for helpful comments on the anti-symmetrization of the Hilbert space. P.A. acknowledges the support received from the Princeton Program in Plasma Science and Technology (PPST). B.L. is funded by the Swiss National Science Foundation through the Postdoc.Mobility Fellowship grant \#P500PT\_211060. H.R. acknowledges the support provided by the U.S. Department of Energy (DOE) through grant DE-FG02-02ER15344.
\end{acknowledgments}

\bibliography{apssamp}% Produces the bibliography via BibTeX.
\end{document}